\documentstyle[aps,prl,epsfig]{revtex}
\begin{document}
\draft

\widetext

\title{Heterogeneous aging in spin glasses}

\author{
Horacio~E.~Castillo$^1$, Claudio~Chamon$^1$,
Leticia~F.~Cugliandolo$^{2,3}$, Malcolm~P.~Kennett$^4$
}
\address{
$^1$ Physics Department, Boston University, Boston, MA 02215, USA \\
$^2$ Laboratoire de Physique Th{\'e}orique de l'{\'E}cole Normale
Sup{\'e}rieure, Paris, France \\ 
$^3$ Laboratoire de Physique
Th{\'e}orique et Hautes Energies, Jussieu, Paris, France \\ 
$^4$ Department
of Physics, Princeton University, Princeton, NJ 08544, USA }

\date{February 16, 2002}

\twocolumn[\hsize\textwidth\columnwidth\hsize\csname@twocolumnfalse\endcsname
\maketitle

\begin{abstract}
  We introduce a set of theoretical ideas that form the basis for an
  analytical framework capable of describing nonequilibrium dynamics in
  glassy systems.  We test the resulting scenario by comparing its
  predictions with numerical simulations of short-range spin glasses. Local
  fluctuations and responses are shown to be connected by a generalized local
  out-of-equilibrium fluctuation-dissipation relation.  Scaling relationships
  are uncovered for the slow evolution of heterogeneities at all time scales.
\end{abstract}
\pacs{PACS: 75.10.Nr, 75.10.Jm, 75.50.Lk, 75.10.Hk}]
\narrowtext

Very slow equilibration and sluggish dynamics are characteristics shared by
disordered spin systems and by other glassy systems such as structural and
polymeric glasses. The origin of this dynamic arrest near and below the glass
transition is currently poorly understood. Studies of the time evolution of
many quantities, such as the remanent magnetization, the dielectric constant,
or the incoherent correlation function, have shown that below the glass
transition the system falls out of equilibrium~\cite{out-of-equil,Eric}. 
This is
evidenced by the presence of aging, {\it i.e.} the dependence of physical
properties on the time since the quench into the glassy state, and also by
the breakdown of the equilibrium relations dictated by the fluctuation
dissipation theorem ({\sc fdt})~\cite{mean-field-dyn,review}.

Most analytical progress in understanding non-equilibrium glassy dynamics
has been achieved in mean-field fully connected spin 
models~\cite{mean-field-dyn},
while numerical simulations have addressed both structural
glasses~\cite{aging-struct} and short-range spin glass
models~\cite{Picco}.  Until recently, however, experimental, numerical,
and analytical studies have mainly focused on global quantities, such as
global correlations and responses, which do not directly probe local
relaxation mechanisms. Local regions that behave differently from the bulk,
or dynamic heterogeneities, could be crucial to understand the full
temporal evolution, and have received considerable
experimental~\cite{heterogeneities,confocal,Nathan} and
numerical~\cite{numerical} attention lately.  However, no clear theoretical
picture has yet emerged to describe the local nonequilibrium dynamics of the
glassy phase.

Here we introduce such a theoretical framework, and test its predictions via
numerical simulations of a short-range spin glass model. 
We show that local correlations and responses are
linked, and we find scaling properties for the heterogeneities that connect
the evolution of the system at different times. This universality may
provide a general basis for a realistic physical understanding of glassy
dynamics in a wide range of systems.

The framework that we propose is motivated by an analogy~\cite{reparam1} between
aging dynamics and the well-known statics of Heisenberg magnets. For
concreteness, we test its predictions against Monte Carlo 
simulations on the prototypical
spin glass model, the three dimensional Edwards-Anderson
(3DEA) model, $H = \sum_{\langle ij \rangle} J_{ij} s_i s_j$,
where $s_i = \pm  1$ and the nearest-neighbor couplings are $J_{ij}=\pm  1$
with equal probability. We argue that two dynamical 
{\it local} quantities, the
coarse-grained local correlation $ C_r(t,t_w) \equiv \frac1{V} \sum_{i\in V_r}
\overline s_i(t) \overline s_i(t_w) $ and integrated response $\chi_r(t,t_w) 
 \equiv \frac1{N_f} \sum_{k=1}^{N_f} \frac1{V} \sum_{i\in V_r} 
 {\overline s_i(t)|_{h^{(k)}}-\overline s_i(t) \over h^{(k)}_i} $
are essential to understand the mechanisms
controlling the dynamics of glassy systems. The spins are represented by
$s_i$ in the absence of an applied field and by $s_i|_{h^{(k)}}$ in the
presence of one.  $\overline s_i(t) \equiv \frac{1}{\tau}
\sum_{t'=t-\tau}^{t'=t-1} s_i(t')$
is the result of coarse-graining the spin over a small time-window
[typically, $\tau=1000$ Monte Carlo steps (MCs)].  $V_r$ is a cubic box with
volume $V$ centered at the point $r$.  By taking $V$ to be the volume of the
whole system, the bulk or global correlation $C(t,t_w)$ and response
$\chi(t,t_w)$ are recovered. 
Two generic times after preparation are
represented by $t_w$ and $t$, with $t_w\leq t$.  When the system is not in
equilibrium, time dependences {\em do not} reduce to a dependence on the time
difference $t-t_w$.  We measure a staggered local integrated linear response by
applying a bimodal random field on each site $h^{(k)}_i=\pm  h$ during the
time interval $[t_w, t]$.  Linear response holds for the values of $h$ that
we use.  The index $k=1,\dots,N_f$ labels different realizations of the
perturbing field. We use random initial conditions.  The thermal histories,
i.e. the sequences of spins and random numbers used in the MC test, are the
same with and without a perturbing field.

\begin{figure}
\vspace{-0.25cm}
\epsfxsize=8.5cm
\epsfbox{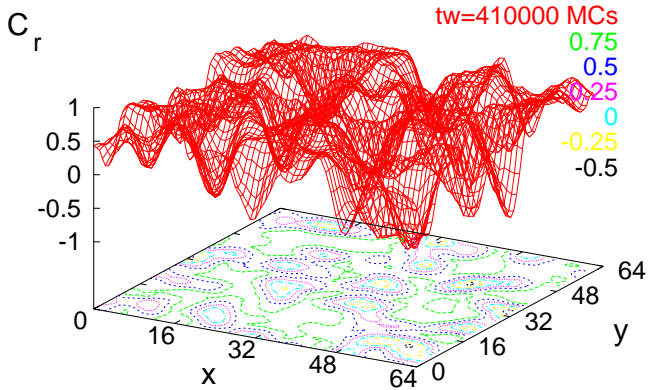}
\vspace{0.05cm}
\caption{The local correlations on a 2D cut of the
cubic cell. The linear size is 
$L=64$. $T = 0.72 \, T_c$, $V=3^3$,  
$t_w= 4.1\times   10^5$ MCs and $t= 2.8 \times   10^6$ MCs.}
\label{fig1}
\end{figure}

In a disordered spin model, the coarse-grained local magnetization typically
vanishes, but the local correlation is non-trivial.  Averaged over disorder
and the thermal history, this correlator defines the Edwards-Anderson
parameter $q_{\sc ea}$ when $t_w\to \infty$, and $t-t_w\to\infty$
subsequently.  
Can we detect the growth of local order~\cite{Fihu} by analyzing the 
evolution of the local correlator, as one easily can
for a system undergoing ferromagnetic domain growth?
In Fig.~\ref{fig1} we show the local
correlation for fixed $t_w$ and $t$ on a 2D cut of the 3DEA model.  Regions
with large values of $C_r$ are intertwined with regions with a small value of
$C_r$ as shown by the contour levels. This behavior persists for all $t_w$
and $t$ that we can reach with the simulation and a more sophisticated
analysis is necessary to identify a growing order in this system.

It is clear from Fig.~\ref{fig1} that different sites have distinct dynamics.
Analysis of the local correlation for fixed $t_w$ as a function of $t$ shows
that in general the relaxation is non-exponential -- this is often
ascribed to the presence of heterogeneous dynamics.  
How can one characterize
the heterogeneous dynamics and determine its origin?  We argue that relevant
quantities are the probability distribution ({\sc pdf}) of the local
correlation, $\rho(C_r(t,t_w))$, the {\sc pdf} of the local integrated
response, $\rho(\chi_r(t,t_w))$, and the joint {\sc pdf}
$\rho(C_r(t,t_w),\chi_r(t,t_w))$ and we start by discussing the latter.  

\begin{figure}
\center 
\vspace{-0.25cm}
\epsfxsize=7.5cm
\epsfbox{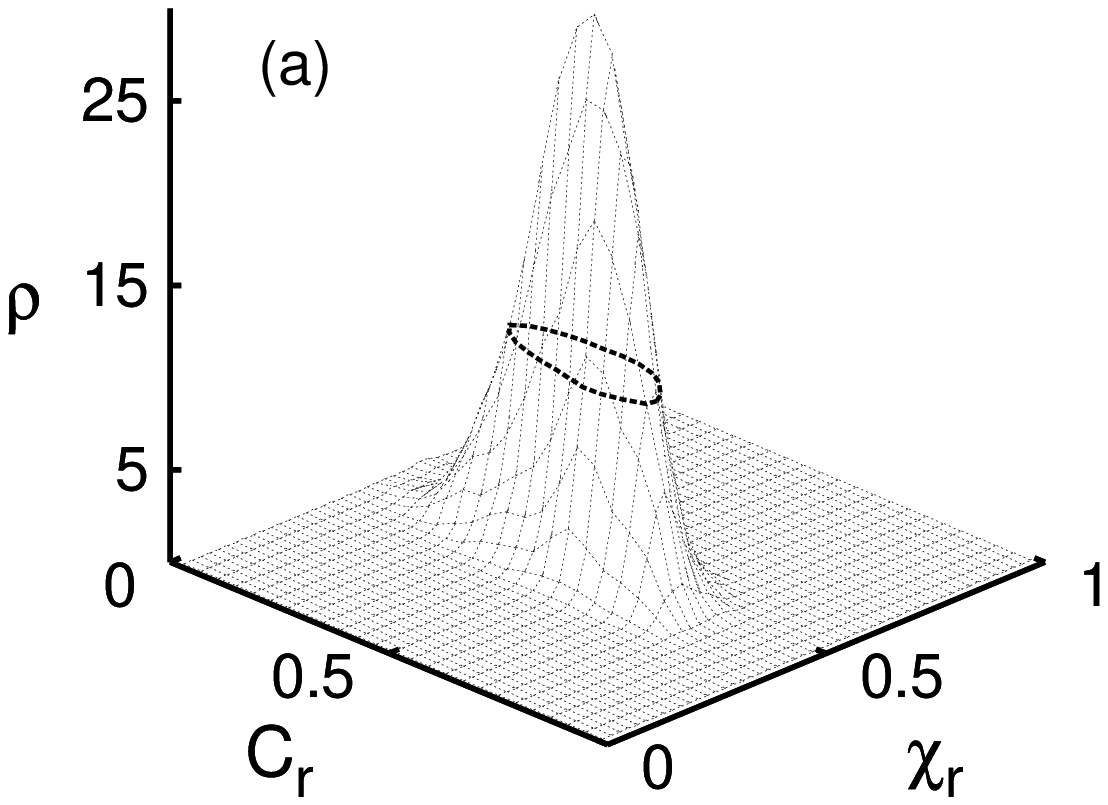}
\epsfxsize=7.5cm
\epsfbox{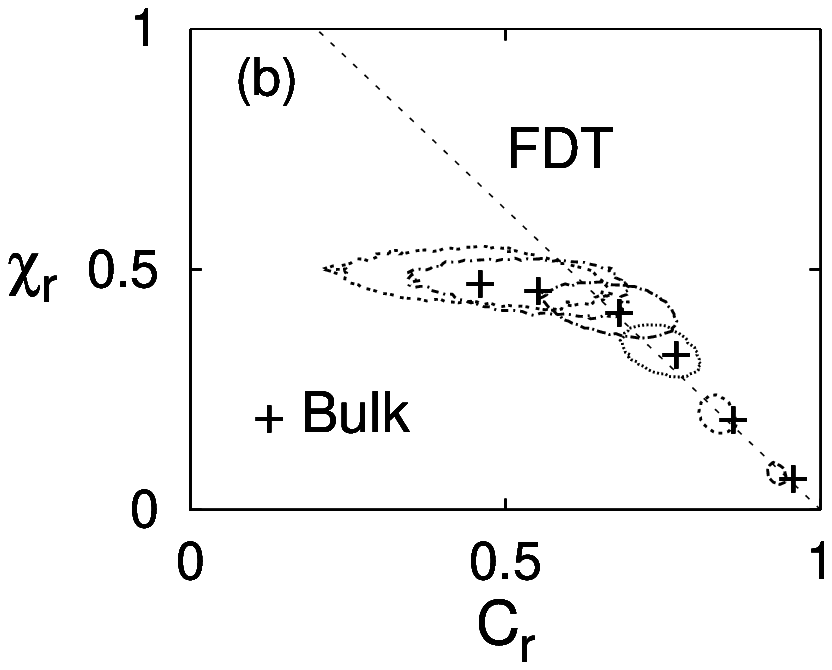}
\caption{
$L=64$, $T = 0.72 \, T_c$, $V = 13^3$, $t_w=4 \times 10^4$.
(a) Surface plot of the joint {\sc pdf}  for $t/t_w=16$.
  Points within the shown contour account for 2/3 of the total probability. (b)
  Time evolution of $\rho(C_r(t,t_w),\chi_r(t,t_w))$, for 
  $t/t_w=1.00005$, $1.001$, $1.06$, $2$, $8$, $32$ (from right to
  left). Each contour contains $2/3$ of the total probability
  for the given pair of times.  The line marked {\sc fdt}
  is the equilibrium relation between $\chi_r$ and $C_r$.  The crosses are
  the bulk $\chi(C)$ for the same pairs of times. }
\label{fig3} 
\end{figure}

The {\sc fdt} relates the
correlation of spontaneous fluctuations to the integrated linear response of
a chosen observable ({\it e.g.} $C_r(t,t_w)$ and $\chi_r(t,t_w)$ averaged over
thermal histories) at equilibrium.  Glassy systems modify the {\sc fdt} in a
particular way first obtained analytically for mean-field
models~\cite{mean-field-dyn}, later verified numerically in a
number of realistic models~\cite{numerics-fdt,Ludo}, and
more recently tested experimentally~\cite{exp-fdt}. A parametric plot
of the bulk integrated response, $\chi(t,t_w)$, against the bulk correlation,
$C(t,t_w)$, for fixed and long waiting-time $t_w$ and using $t$ as a
parameter, approaches a non-trivial limit, $\chi(C)$, represented by the
crosses in Fig.~\ref{fig3}(b). The curve has a
straight section, for which $t-t_w\ll  t_w$ and the correlation decays from
$1$ to $q_{\sc ea}$ with a slope $-1/T$ as found with the equilibrium {\sc
fdt}.  Beyond this point, as $t$ increases towards infinity, the curve
separates from the {\sc fdt} line.  Now, consider each lattice site for fixed
times $t_w$ and $t$.  
 If we plot points for the pairs $(C_r(t,t_w),
 \chi_r(t,t_w))$, where will they lie?

When $t_w\to\infty$ and $t-t_w\ll  t_w$, all local correlators satisfy the
{\sc fdt} strictly once averaged over thermal histories, since the magnitude
of local deviations from the {\sc fdt} has an upper bound~\cite{Cudeku}.  We
have checked that $C_r(t,t_w)$ and $\chi_r(t,t_w)$ obey the {\sc fdt} for an
individual thermal history apart from small fluctuations 
[see Fig.~\ref{fig3}(b)].

For the regime of widely separated times we propose an analysis
similar in spirit to the one that applies to the low energy
excitations of the Heisenberg model.  
There,
the free energy for the
coarse grained magnetization ${\vec m(\vec r)}$ is
$
F = \int d^d r [ ({\vec \nabla}_{\vec r} {\vec m}({\vec r}))^2 
+ V(|{\vec m}({\vec r})|) ]
$.
A spontaneous symmetry breaking signals the
transition into the ordered phase $\langle \vec m \rangle = \vec m_0 \neq 0$,
in which the order parameter has both a uniform length (the radius of the
bottom of the effective potential $V(|\vec m|)$), and a uniform direction.  
$F$ is
invariant under uniform rotations ${\vec m(\vec r)} \rightarrow {\cal R}
{\vec m(\vec r)}$. The lowest energy excitations (spin waves) are obtained
from the ground state by leaving the length of the vector invariant and
applying a slowly varying rotation to it: $\vec m(\vec r) = {\cal R}(\vec r)
\vec m_0 $. These are massless transverse fluctuations (Goldstone modes). In
contrast, longitudinal fluctuations, which change the magnitude of the
magnetization vector, are massive and energetically costly.

Let us now apply the same kind of analysis to the dynamics of the spin
glass. Here, the relevant fluctuating quantities are the coarse grained local
correlations $C_r$ and their associated local integrated responses
$\chi_r$.  In Ref.~\cite{reparam1} we derived an effective action for
these functions that becomes invariant under a global time-reparametrization
$t \to h(t)$ in the aging regime.  This symmetry leaves the bulk relation,
$\chi(C)$, invariant.  A uniform reparametrization is analogous to a global
rotation in the Heisenberg magnet, and the curve $\chi(C)$ is analogous to
the surface where $V(|\vec m|)$ is minimized.  Hence, we expect that for fixed
long times $t_w$ and $t$ in the aging regime, 
the local fluctuations in $C_r$ and $\chi_r$ should
be given by smooth spatial variations in the time reparametrization,
$h_r(t)$, i.e. $C_r(t,t_w) = C_{\sc sp}(h_r(t),h_r(t_w)) 
\approx C(h_r(t),h_r(t_w))$ where $C_{\sc sp}$ is the global
correlation at the saddle-point level 
that in the numerical studies we approximate by the 
actual global correlation $C$, 
and similarly for $\chi_r$.
These transverse
fluctuations are soft Goldstone modes. Longitudinal fluctuations, which move
away from the $\chi(C)$ curve, are massive and penalized.
This implies the first testable prediction of our theoretical framework:
the pairs $(C_r,\chi_r)$ should follow the curve $\chi(C)$ for the bulk
integrated response against the bulk correlation. 

In Fig.~\ref{fig3}, we test this prediction by plotting the distribution of
pairs $(C_r,\chi_r)$. We find, as expected, that for long times the
dispersion in the longitudinal direction (i.e. away from the bulk $\chi(C)$
curve) is much weaker than in the transverse direction (i.e. along the bulk
$\chi(C)$ curve).
In the coarse grained aging limit we expect the former to
 disappear while the latter should remain.  (This limit corresponds to the
 way actual measurements are performed: the thermodynamic limit is taken
 first to eliminate finite size effects and undesired equilibration; then the
 large $t_w$ limit is taken to reach the asymptotic
 regime; finally, the limit $V\to\infty$ serves to eliminate fluctuations
 through the coarse graining process; in the figure we used a large volume 
 $V=13^3$ to approach the latter limit though we found a 
 similar qualitative behavior for smaller $V$.) 
Figure~\ref{fig3}(a) displays the joint {\sc pdf} $\rho(C_r, \chi_r)$ for
a pair of times $(t_w,t)$ that are far away from each other.
Figure~\ref{fig3}(b) shows the projection of a set of contour levels 
for $t_w$ fixed and six values of $t$.
Even though the data for each contour corresponds to a single
pair of times $(t_w,t)$, the fluctuations span a range of values that, for
the bulk quantities, would require a whole family of pairs
$(t_w,t)$. This reveals that the aging process is non-uniform across a
finite-range model.  

We now turn our attention to a more detailed examination of the time
dependences in the local dynamics. 
A
good $t/t_w$ scaling that breaks down only for very large values of the
subsequent time $t-t_w$ has been obtained for the bulk thermoremanent
magnetization experimentally~\cite{Eric} and for the bulk correlation
numerically~\cite{Picco} once the stationary ($ t-t_w \ll  t_w$) part of the
relaxation is subtracted, as suggested by the solution to mean-field
models~\cite{mean-field-dyn}.  
For systems that display this particular dependence on $t/t_w$ for the bulk
correlator, a second prediction can be extracted from our
theoretical framework: the distribution $\rho(C_r(t,t_w))$ should
only depend on the ratio $t/t_w$. Even further, if the bulk correlator has
a simple power law form 
$C_{\sc sp}(t,t_w) \sim q_{\sc ea} (t/t_w)^{-\rho}$, an
approximate treatment of fluctuations leads to a rescaling and
collapse of $\rho(C_r(t,t_w))$ even for pairs of times with {\em different}
ratios $t/t_w$.  

Since we are dealing with ratios of times, it is convenient to define 
$h_r(t)=e^{\varphi_r(t)}$, so that 
$C_r(t,t_w)=C_{\sc
sp}(h_r(t)/h_r(t_w))=C_{\sc sp}(e^{\;\varphi_r(t)-\varphi_r(t_w)})$.
Therefore the statistics of local correlations are determined from the
statistical distribution of distances between two ``surfaces'',
$\varphi_r(t)-\varphi_r(t_w)$. In this form, a dynamic theory of short-range
spin glasses is not different from a theory of fluctuating geometries or
elasticity. We propose a simple reparametrization invariant 
effective action for $\varphi_r(t) =
\ln t + \delta\varphi_r(t)$, expanding around $\delta\varphi_r(t) = 0$, with
no zeroth or first order term in $\delta\varphi_r(t)$. 
We assure that the effective action is
reparametrization invariant by taking one time derivative for each time
variable. Thus~\cite{preparation}
$$S=\frac{q_{\sc ea} \rho}{2}\! \int \!\! d^{d}r \!
\int_0^\infty \!\!\!\!\!\! dt 
\int_0^\infty \!\!\!\!\!\! dt^\prime
\;\nabla \dot{\varphi}_r(t) 
\;\nabla \dot{\varphi}_r(t^\prime)
\; e^{-\rho|\varphi_r(t) - \varphi_r(t^\prime)|}.$$
where the last factor penalizes fast time variations of $\varphi_r$
and the $\nabla$ ensure that spatial variations are smooth. 
Expanding to lowest order in $\delta\varphi_r(t)$ yields $\varphi_r(t) -
\varphi_r(t_w) = \ln(t/t_w) + \delta\varphi_r(t) - \delta\varphi_r(t_w)
\simeq \ln(t/t_w) + (a + b\ln(t/t_w))^\alpha X_r(t,t_w),$ where $a$ and $b$ are determined by the magnitude of the fluctuations, and
$X_r(t,t_w)$ is a random variable drawn from a time-independent {\sc
pdf} that governs the fluctuations of the surfaces. In our 
approximation, which describes 
uncorrelated drift between two surfaces (i.e. a random walk), 
$\alpha = 1/2$.

\begin{figure} 
\epsfxsize=3.2in
\center
\vspace{-0.4cm}
\epsfbox{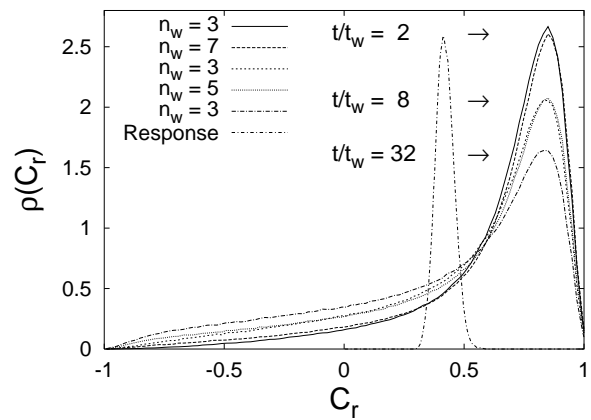} 
\caption{
{\sc pdf} of the local correlations
for
the ratios $t/t_w = 2$, $8$, and $32$, averaged over $36$ realizations
of disorder. $L=32$,  
$T = 0.72 \, T_c$ and  $V=3^3$.
The waiting-times are $t_w=10^4\, (2^{n_w-1})$ MCs, 
with $n_w$ given
in the key. We include the {\sc pdf} 
of the local integrated responses, $\chi_r(t,t_w)$, for 
$n_w=3$ and $t/t_w=2$. 
($\rho(\chi_r)$ is divided by $4$ to fit on the figure.)}
\label{fig2} 
\end{figure}

Figure~\ref{fig2} displays $\rho(C_r(t,t_w))$ for several choices of the
ratio $t/t_w$.  
Interestingly enough, all the curves have a
noticeable peak at a value of $C_r$ that is independent of $t$ and $t_w$,
with a height that decreases significantly with increasing ratio $t/t_w$.  
The form derived above for $\varphi_r(t) -
\varphi_r(t_w)$ explains the approximate collapse of 
$\rho(C_r(t,t_w))$ for a
fixed ratio $t/t_w$, as shown in Fig.~\ref{fig2} for a small value of $V$. 
(Due to mixing with the stationary part, the $t/t_w$
scaling worsens when $V$ increases.)  Barely noticeable in Fig.~\ref{fig2} is
a slow drift of the curves for increasing values of $t_w$ at fixed ratio
$t/t_w$ such that the height of the peak decreases while the area below the
tail at lower values of $C_r$ increases. This trend leads to the
``sub-aging'' scaling observed for bulk quantities~\cite{Eric,Picco}.

\begin{figure}
\center
\epsfxsize=3.1in
\vspace{-0.25cm}
\epsfbox{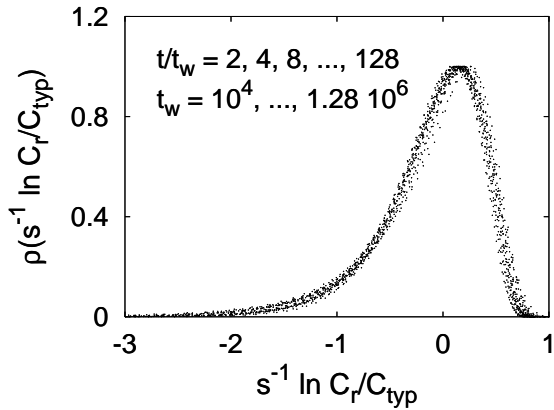}
\vspace{0.1cm}
\caption{ Scaling of $\rho(C_r(t,t_w))$. 
$L = 32$, $T = 0.72 \, T_c$, $V=13^3$, and $28$ pairs
$(t,t_w)$, with $t=10^4\, (2^{n-1})$ MCs, $n=1,\ldots,8$ and
$t_w=10^4\, (2^{n_w-1})$ MCs, $n_w=1,\ldots,n-1$. 
$C_{\rm typ} \equiv \exp \langle
\ln C \rangle$, where $\langle \cdots \rangle$ denotes an average 
over the distribution and $s$ rescales the maximum value of each curve.
With this choice of $V$ we eliminate the $C_r<0$, cfr. 
Fig.~\ref{fig2}.}
\label{fig5} 
\end{figure}

Furthermore, the above expression for
$\varphi_r(t)-\varphi_r(t_w)$ implies that 
the {\sc pdf}s for all of the $28$ pairs of times $(t,t_w)$
 should collapse by rescaling with two parameters: $\ln C_{typ}$ and $s$,
corresponding respectively to the nonrandom part in
$\varphi_r(t)-\varphi_r(t_w)$ and to the width of the random part
(see Fig.~\ref{fig5}). The
scaling curve itself gives the {\sc pdf} for $X_r(t,t_w)$. 
The rather good collapse of the curves  should
be improved by further knowledge of $C_{\sc sp}$.

A local separation of time-scales leads to a reparametrization 
invariant action~\cite{reparam1} with a soft mode that controls
the aging dynamics. Our framework, based on the analogy with Heisenberg 
magnets, predicts the existence of a local relationship
between $C_r$ and $\chi_r$, expressed by $\rho(C_r(t,t_w),\chi_r(t,t_w))$
being sharply concentrated along the global $\chi(C)$ 
in the $(C,\chi)$ plane. Under
additional assumptions, we obtained 
the scaling behavior of $\rho(C_r(t,t_w))$ for all
$(t,t_w)$. Our simulations both confirm these predictions and uncover
striking regularities in the geometry of local 
fluctuations~\cite{preparation}.  
These results
open the way to a systematic study of local dynamic fluctuations in glassy
systems and they suggest a number of exciting avenues for future research.
On the theoretical side, this framework could be applied to a large variety
of glassy models, including those without explicit disorder.  More
interesting still are the experimental tests suggested by this work.  For
instance, the local correlations of colloidal glasses are accessible
experimentally with the confocal microscopy technique~\cite{confocal}.
Similarly, cantilever measurements of noise spectra~\cite{Nathan} allow
probing of local fluctuations in the glassy phase of polymer melts.  These
are just two examples: any experiment that measures local fluctuations in
glassy systems is a potential candidate for testing our ideas.

We thank D. Huse and J. Kurchan for useful discussions. 
Supported in part by the NSF (grant DMR-98-76208) and
the Alfred P. Sloan Foundation. Supercomputing time was 
allocated by the Boston University SCF.


\vspace{-0.5cm}

\begin{thebibliography}{99}

\vspace{-1.5cm}

\bibitem{out-of-equil}
L. C. E. Struick, {\it Physical aging in amorphous polymers and 
other materials} (Elsevier, 1978).

\bibitem{Eric} E. Vincent {\em et al.}, 
in {\it Proceedings of the Sitges conference} (E. Rubi ed.,
Springer-Verlag, 1997).

\bibitem{mean-field-dyn}  
L. F. Cugliandolo and J. Kurchan, 
Phys. Rev. Lett. {\bf 71}, 173 (1993);
L. F. Cugliandolo and J. Kurchan, 
J. Phys. A {\bf 27},  5749 (1994). 

\bibitem{review} J-P Bouchaud {\it et al}, in {\it Spin glasses and 
random fields} A. P. Young ed (World Scientific, 1998). 

\bibitem{aging-struct}
J-L Barrat and W. Kob, Eur. Phys. J. B {\bf 13}, 319 (2000).

\bibitem{Picco} M. Picco, F. Ricci-Tersenghi, F. Ritort, 
Eur. Phys. J. B {\bf 21}, 211 (2001).

\bibitem{heterogeneities} 
M. D. Ediger, Annu. Rev. Phys. Chem. {\bf 51}, 99 (2000).

\bibitem{confocal} A. van Blaaderen and  P. Wiltzius, 
Science {\bf 270}, 1177 (1995); 
E. R. Weeks {\em et al.}, 
Science {\bf 287}, 627 (2000);
W. K. Kegel and A. van Blaaderen, 
Science {\bf 287}, 290 (2000).

\bibitem{Nathan}
E. Vidal-Russell and N. E. Israeloff, 
Nature {\bf 408}, 695 (2000).  

\bibitem{numerical}
P. H. Poole {\it et al}
Phys. Rev. Lett. {\bf 78}, 3394 (1997);
A. Barrat and  R. Zecchina,  
Phys. Rev. E {\bf 59} R1299 (1999);
F. Ricci-Tersenghi and R. Zecchina, 
Phys. Rev. E {\bf 62},
R7567 (2000);
C. Bennemann {\it et al} 
Nature, {\bf 399}, 246 (1999);
W. Kob {\em et al.}, 
Phys. Rev. Lett. {\bf 79}, 2827 (1997).
              
\bibitem{reparam1} C. Chamon {\it et al}, cond-mat/0109150.

\bibitem{Fihu} D. S. Fisher and D. A. Huse, 
Phys. Rev. Lett. {\bf 56},  1601 (1986).

\bibitem{numerics-fdt} 
S. Franz and H. Rieger, J. Stat. Phys. {\bf 79}, 749 (1995).
E. Marinari {\it et al}
J. Phys. A {\bf 33}, 2373 (2000);
W. Kob and J-L. Barrat,  
Eur. Phys. J. B {\bf 13}, 319 (2000);
A. Barrat {\it et al}, Phys. Rev. Lett. {\bf 85}, 5034 (2000); 
H. Makse and J. Kurchan, Nature {\bf 415}, 614 (2002);
J-L. Barrat and L. Berthier, cond-mat/0110257;
Phys. Rev. E {\bf 57}, 3629 (1998).

\bibitem{Ludo} A. Barrat and L. Berthier, Phys. Rev. Lett. 
{\bf 87}, 087204 (2001).

\bibitem{exp-fdt} T. S. Grigera and N. E. Israeloff, 
Phys. Rev. Lett. {\bf 83}, 5038 (2000). 
L. Bellon, S. Ciliberto, C. Laroche, 
Europhys. Lett. {\bf 53}, 511 (2001).
D. Herisson and M. Ocio, cond-mat/0112378.

\bibitem{Cudeku} L. F. Cugliandolo, D. S. Dean, J. Kurchan, 
Phys. Rev. Lett. {\bf 79}, 2168 (1997). 

\bibitem{preparation} H. E. Castillo, 
C. Chamon, L. F. Cugliandolo, M. P. Kennett, in preparation.

\end{thebibliography}
\end{document}